# Current assessment of the Red Rectangle band problem

R. J. Glinski • P. D. Michaels • C. M. Anderson • T. W. Schmidt •
R. G. Sharp • M. L. Sitko • L. S. Bernstein • H. Van Winckel

**Abstract**  In this paper we discuss our insights into several key problems in the identification of the Red Rectangle Bands (RRBs).  We have combined three independent sets of observations in order to try to define the constraints guiding the bands.  We provide a summary of the general behavior of the bands and review the evidence for a molecular origin of the bands.  The extent, composition, and possible absorption effects of the bands are discussed.  Comparison spectra of the strongest band obtained at three different spectral resolutions suggests that an intrinsic line width of individual rotational lines can be deduced. Spectroscopic models of several relatively simple molecules were examined in order to investigate where the current data are weak.  Suggestions are made for future studies to enhance our understanding of these enigmatic bands.

R. J. Glinski & P. D. Michaels
Department of Chemistry, Tennessee Tech University, Cookeville, TN  38505, USA
e-mail: rglinski@tntech.edu

C. M. Anderson
Department of Astronomy, University of Wisconsin, Madison, WI  53706, USA

T. W. Schmidt
School of Chemistry, Building F11, University of Sydney, NSW, 2006, Australia
e-mail: T.Schmidt@chem.usyd.edu.au

R. G. Sharp
Anglo-Australian Observatory, PO Box 296, Epping, NSW, 1710, Australia
e-mail: rgs@aao.gov.au

M. L. Sitko
Department of Physics, University of Cincinnati, Cincinnati, OH  45221, USA
e-mail: sitkoml@ucmail.uc.edu

L. S. Bernstein
Spectral Sciences, Inc., Burlington, MA  01803, USA
e-mail: larry@spectral.com

H. Van Winckel
Instituut voor Sterrenkunde, Katholieke Universiteit Leuven, Celestijnenlaan 200D, 3001 Leuven, Belgium
e-mail: hans.vanwinckel@ster.kuleuven.be

# 1 Introduction

The Red Rectangle is a biconical nebula centered around the post-AGB star, HD 44179 (Cohen et al. 1975), and an as yet uncharacterized companion (Waelkens et al. 1996; Men'shchikov et al. 2002; Sitko et al. 2008), which are evolving on a binary track (Van Winckel 2007). An opaque disc/torus complex of molecules, dust, and macroscopic grains surrounds the binary (Waters 1998; Jura 1997). In addition to its spectacular morphology (Cohen et al. 2004), the nebula is special as a strong source of various molecular emission features in the ultraviolet (Reese and Sitko 1996; Glinski et al. 1997; Sitko et al. 2008), visible (Warren-Smith et al. 1981; Schmidt and Witt 1991), and infrared (Russell et al. 1978; Sloan et al. 1993; Hora et al. 1999; Song et al. 2003). The object is also one of the strongest known sources of the extended red emission (ERE) (Witt and Boroson 1990; Van Winckel et al. 2002). Recently, Vijh and co-workers (2005; 2006) have characterized a blue luminescence that is thought to derive from emission from a distribution of small neutral PAHs. Hobbs at al. (2004) have presented their high-resolution studies of the lines from atoms and diatomic molecules on the central source.

We are interested in the as yet unidentified groups of narrow emission features, which lie upon the ERE between about 5500 and 7000 Å. First observed in detail by Schmidt et al. (1980), several other groups have sought to characterize these Red Rectangle bands (RRBs) (Warren-Smith et al. 1981; Schmidt & Witt 1991; Scarrott et al. 1992; Sarre et al. 1995). Schmidt & Witt (1991) demonstrated that the intensities of the RRBs are sharply peaked at the walls of the biconical outflow cavities. Van Winckel et al. (2002), presented a high signal-to-noise, medium resolution study of all of the features lying upon the ERE and listed an extensive catalogue of bands. Sharp et al. (2006) reported a spectroscopic modeling study of the dominant 5799 Å RRB based on their extensive higher resolution IFU observations across the nebula.

In an earlier paper, Glinski and Anderson (2002) examined the band profiles and positions of the peak maxima of the 5799 Å band using the IFU Spectrometer on the WIYN telescope. That paper drew the following conclusions from the data: (1) At offsets from the central source greater than about 8 arcsec, spectra obtained at any two points in the nebula, equidistant to the central source, were entirely congruent; this suggested that a photostationary-state mechanism was responsible for the rotational energy distribution of the emitter. (2) Bands observed within 8 arcsec showed evidence of radiative transfer, i. e. absorption effects, as indicated by north/south differences in the band profiles. (3) The peak maximum of the 5799 Å band did not shift to the blue of that wavelength out to offsets of 17 arcsec; in fact, the peak maximum shifted slightly to the red at wide separations, which could be understood in terms of a hypothetical model spectrum.

In this paper, we report inferences made by combining data from three previously reported sets of observations together with newer observations using the WIYN fiber array spectrometer, which has been modified to improve the spatial resolution. In an effort to provide guidance for future observational and spectroscopic modeling studies, we focus on the following questions: Are the absorption effects real and can they be unambiguously defined? What is the intrinsic line-width that must be applied to each rovibronic line in the modeling? And, what is the extent of each RRB and how many RRBs can be assigned to the same emitter? Finally, we make suggestions for future observations that would address these questions.

## 2 Observations

Most of the observations discussed here have been described in detail elsewhere. Table 1 presents a summary of these observations. The latest two WIYN observations have not been described previously; the parameters of these observations are the same as those in 2001 with the exception that the Cassegrain Instrument Adaptor System (CassIAS) was used. The CassIAS configuration places the 91-fiber DENSPAK fiber array at the Cassegrain focus resulting in an improved spatial resolution shown in the Table. In the latest observations, the central source was placed at the bottom of the array with the north-east cone wall running up the center of the array.

**Table 1** Summary of observations.

| Date | Instrument | Configuration | Spatial Resolution | R at 5800 Å | Notes |
|---|---|---|---|---|---|
| 2002[a] | NNT La Silla | long slit, 1″ wide | 0″.268/pix | 5,500 | |
| 2005[b] | VLT-UT2 Paranal | Argus IFU 125 x 80 pixels 37″ x 28″ | 0″.30/pix | 37,000 | 5590-5830Å |
| 2001[c] | WIYN Kitt Peak | DENSPAK 13 x 7 fibers 43″ x 27″ | 3″ fibers on 4″ centers | 15,000 | 7 Mar |
| 2002 | WIYN | CassIAS[d] 13 x 7 fibers 21″.5 x 13″.5 | 1″.5 fibers on 2″ centers | 15,000 | 2&3 Nov 6600Å data |
| 2004 | WIYN | CassIAS[d] | 1″.5 on 2″ centers | 15,000 | 27&28 Feb |

[a]Van Winckel et al. 2002
[b]Sharp et al. 2006; the VLT spectra shown herein were constructed by averaging a nine-pixel square centered on the stated position.
[c]Glinski & Anderson 2002
[d]see text

## 3  General trends in the bands

The general trends in the 5799 and 5853 Å bands are shown in Figure 1.  The figure provides a graphical summary of the behavior described by Schmidt and Witt (1991); Scarrott et al. (1992); and Van Winckel et al. (2002).  There are several clear trends in the 5799 Å band at the four relatively wide offsets from the star: while there is a small variation in the position of the peak maximum, the band broadens in a regular way as the nebula is sampled closer to the star.  Although the broadening appears to be an integral part of the band, it is important for spectroscopic modeling to establish how much of the flux in the "band" belongs to one emitter.  For instance, is the small shoulder at 5795 Å or the broad shoulder near 5807 Å part of the 5799 Å band?  This is addressed further in Section 7, below.

In all of the spectra reported thus far, the 5799 Å band was found to decrease in integrated intensity *monotonically* from near the center to the farthest detectable reaches, as its shape changed in a regular way.  We note, however, that there is an inherent minimum spatial resolution due to the average terrestrial "good seeing" of about 0.8 arcsec.  This means that the ground-based observations may be smoothing out any knot or ladder-rung structure in these bands as is seen in the broad-band HST images (Cohen et al. 2004).  It is known that the RRBs correlate with the ERE in its coarse structure (Schmidt and Witt 1991; Kerr et al. 1999); but it is not yet known whether the RRBs correlate with the fine spatial structure.

3.1  On the validity of the RRB/DIB correspondence hypothesis

It can be seen that the position of the 5797 Å diffuse interstellar band (DIB) falls only at the bluest extreme of the 5799 Å band.  As discussed previously (Van Winckel et al. 2002; Glinski and Anderson 2002), there does not seem to be a clear convergence between the RRB and the DIB as had been suggested (Fossey 1991; Sarre 1991, Scarrott et al. 1992; Sarre et al. 1995).  Indeed, we do not see clear evidence for the link in spectra published subsequently; instead, there is further evidence that the position of the peak maximum turns slightly back to the red at large offsets (Sharp et al. 2006).  This is in conflict with the hypothesis that these RRBs and the corresponding DIBs share the same carrier; although there may still be some validity to the idea if somehow the lower states resulting from the RRB emission are different from those in the DIB absorption (Sarre 2006).

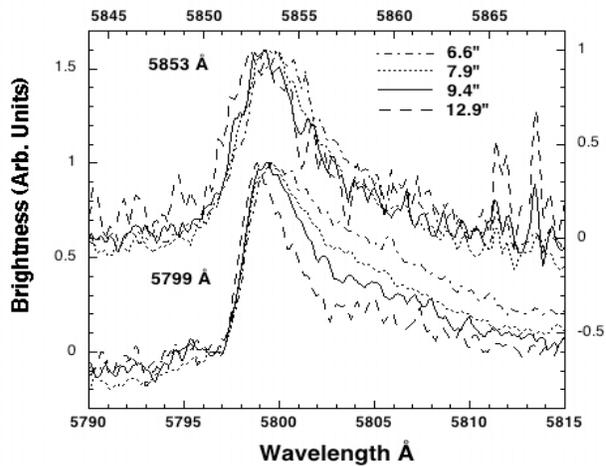

**Fig. 1** Normalized WIYN fiber spectra of the 5799 and 5853 Å bands at 4 offsets along the NE spike. (We feel that the 5853 Å band at 12.9 arcsec is too noisy to indicate a trend.)

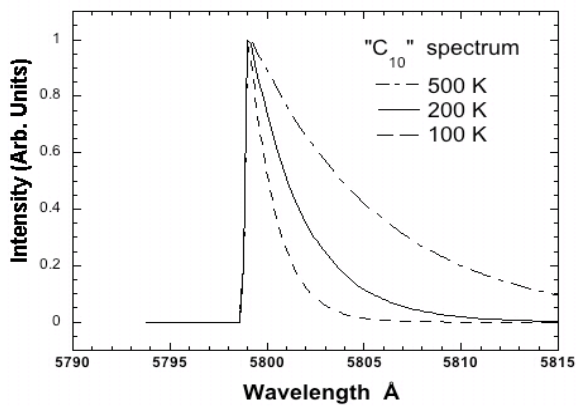

**Fig. 2** Calculated spectrum of a moderately large molecule as an emission spectrum between singlet state at three different temperatures, arbitrarily placed at 5799 Å. The model uses $B´ = 0.0095$ and $B´´ = 0.010$ cm$^{-1}$ (a 3% elongation of the bonds) and an intrinsic line width of 0.1 Å.

## 4 Molecular behavior of the bands

Emission by molecules, from small (Bennett 1987; Warren-Smith et al. 1981) to large (Sarre et al. 1995; Rouan et al. 1997; Le Coupanec et al. 1999;), as the origin of these bands has been discussed previously. Quantitative modeling has been applied to the spectroscopic analysis from three different points of view: rotational band contour modeling (Scarrott et al. 1992; Rouan et al. 1997; Glinski and Anderson 2002), proposed vibrational sequence structure (Sharp et al. 2006), and an overall vibrational analysis (Glinski & Nuth 1997). Yet there is continued ambiguity as to the identity of the emitter. The profiles of the red-shaded RRBs are highly suggestive of rotational contours of a molecule at various temperatures. This is illustrated in Figure 2, for a generic, medium-mass, pseudo-diatomic molecule having the approximate moment of inertia of a $C_{10}$ linear molecule. Since the rotational populations are likely maintained by photoexcitation processes, ultimately spectroscopic modeling will have to include non-local thermodynamic equilibrium energy distributions. Detailed discussions of unassigned optical spectra of molecules in space are given by Schmidt and Sharp (2005) and by Sarre (2006a).

It can be seen by comparison of Figures 1 and 2 that the broadening of a rotational profile with excitation temperature is consistent with the RRB profile trend. This generic modeling is meant to show that filling the central part of the band is readily accomplished with a molecule of medium size. We have found that models of the prolate rotor, tetracene, or the oblate rotor, coronene, will yield a similar figure, with slight differences in the way the peak's shape and maximum evolves. These two simple PAHs have similar rotational constants to both $C_{10}$ and $C_{60}$, $B_{rot} \approx 0.01$ cm$^{-1}$. Three features of the RRB are not well fit by the generic models, however: the peak maximum rounds off and shifts strongly to the red in spectra obtained close to the star; the blue edge is not filled in; and the shoulder at about 5805 Å is also not filled in.

Sharp et al. (2006) have addressed these issues by proposing that a sequence structure occurs in the origin band of a larger, symmetric PAH molecule as suggested by Sarre et al. (1995). Reasonable assumptions have allowed satisfactory fits of all three features, all at once.

In this paper, we examine the three band features separately. We suggest that the three aspects may be the result of different, and themselves important, effects. We will discuss the following three possibilities: (1) The close-in behavior of the bands is influenced by the effects of optical thickness, accounting for some of the shift to the red of the band maximum. (2) The intrinsic line profile itself may be broadened by velocity dispersions within the nebula, which would fill in the blue edge of the band. (3) Overlapping bands from different weak emitters may distort the profiles. We seek to neither confirm nor deny the viability of previous modeling efforts, only to explore ambiguities in the data.

## 5  Are there optically thick regions?

In a previous paper, Glinski and Anderson (2002) presented evidence that the nebula may become optically thick with respect to the 5799 Å band as one samples closer to the star.  There was evidence of there being more absorbing material in southern lines-of-site than in northern lines-of-site equidistant from and within 8 arcsec of the star.  To account for this, there only needed to be 10 percent more absorbing molecules in the southern column and its oscillator strength needed only be about 0.01.  These effects were consistent with the observation that the molecular torus is optically thick with respect to the 11.3 μm "PAH" band inside about 5 arcsec (Bregman et al. 1993).  It was our view that the distribution of the absorbing material of the RRB was approximately toroidal (and extended), so we carefully examined the band profiles as we crossed one of the spikes from inside the cone wall toward the dark outer regions.  In the newer WIYN spectra, we find several examples of significant differences in band profile in two fibers near a spike (inside and outside the cone walls).  In Figure 3 we show one example of the differences seen in adjacent fibers, one on the spike and one inside the cone wall; these fibers are equidistant from the central star (within about 0.2 arcsec) and are centered only 2 arcsec apart.  Similar effects are observed in unpublished VLT-IFU spectra across other cone walls (Sharp et al. 2006).

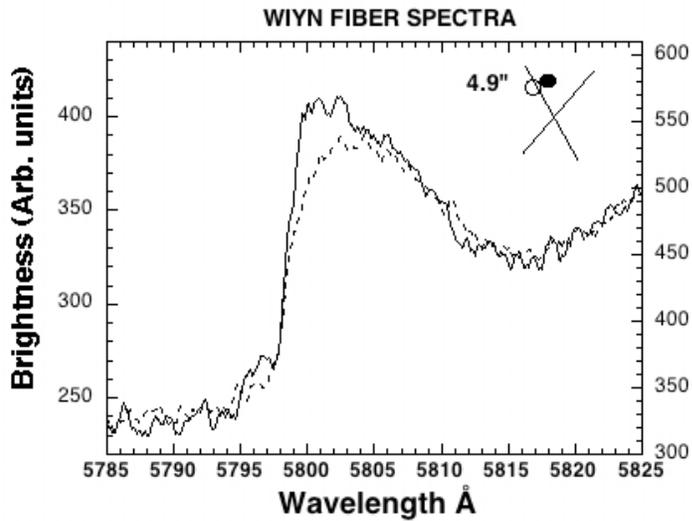

**Fig. 3** Spectra obtained in two fibers equidistant from the star near the NE spike. The spectrum on the spike (dotted line corresponds to the position of the open circle) would seem to reflect a hotter rotational temperature, but we suggest that this can be accounted for by an absorption effect.

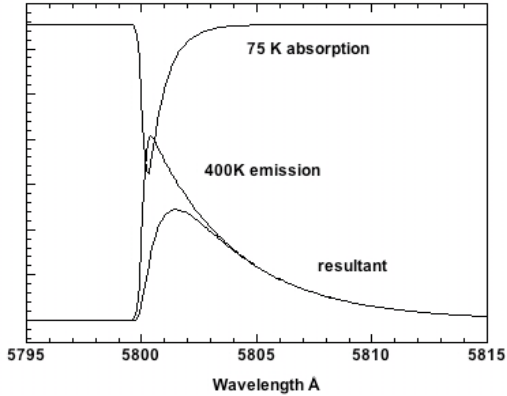

**Fig. 4:** How colder absorbing material could effect the emission profile. The model is similar to the one in Fig. 2 for a molecule approximately the size of coronene.

It would appear at first look that the spectrum obtained on the spike corresponds to material with a higher rotational excitation temperature. But we think we can account for the differences seen in Figure 3 by invoking an excess of absorbing material in the line of site of the fiber on the spike over that of the fiber inside the cone. It would be reasonable that the absorbing material would have a colder excitation temperature since it is shielded from the starlight.

Figure 4 shows how the absorption effect would work to dramatically change an emission profile in model spectra. A column of colder absorbing material can chop out part of a band near the head and give it the appearance of a much hotter distribution.

We believe this could explain several trends that appear in the data: the abrupt take-off to the red of the peak maximum when observed within about 7 arcsec (Glinski and Anderson 2002; Van Winckel et al. 2002) and the distinct rounding-off of the bands within about 5 arcsec (Van Winckel et al. 2002; Sharp et al. 2006). We offer these as caveats before trying to make spectroscopic fits of the bands. Fitting of pure emission bands to the RR spectra inside about 8 arcsec may not adequately represent the observed contours. We believe further analysis needs to be done to sort out the behavior inside of 8 arcsec before "emission-only" modeling is undertaken there. We also note that there is evidence of both atomic and molecular lines that are optically thick, at least near the star. We see evidence of Na I being self-absorbed at several arcsec from the star in the WIYN spectra. Hobbs et al. (2004) show a Ca II line that is self-absorbed on the star. Additionally, the C I line at 1931 Å and the Fourth Positive bands of CO are dramatically self-absorbed near the star (Glinski et al. 1997; Sitko et al. 2008), although it is not clear where the absorbing material is in the nebula.

## 6  What is the intrinsic width of the lines?

Spectroscopic fitting of the observed emission bands begins with calculating the term energies of rotational states in the upper and lower electronic states of a candidate molecule. The appropriate spectroscopic connections are then made between the states to calculate the positions of all the lines (see Fig. 5 in Glinski and Anderson 2002). After this, a Gaussian function of appropriate width is applied to each line and all of these Gaussians are summed to obtain the final contour. The FWHM of the Gaussian is the *intrinsic width* of an individual rotational line; it essentially corresponds to the velocity dispersion in the gas in the aperture along the line of sight. (Although a Gaussian is the simplest function, it is not the only possible model of velocity dispersions. It is assumed they are not Lorentzian, as may be the *natural* line shape.) Therefore, to try to deduce the intrinsic width, we examined the band profiles at wide projected distances from the star where they appear narrowest. We found that we could detect little change in shape in the blue edge of the band beyond about 12 arcsec. Additionally, the band profile does not sharpen with increasing instrument spectral resolution and the blue edge could be well fit to a single Gaussian as shown in Figure 5. We, therefore, suggest that the width of the intrinsic profile is greater than the instrumental resolution of the higher resolution measurements and perhaps as wide as ~1 Å. This gives us a simple hypothesis to pursue, however it does not preclude the possibility that the form of the blue edge may be controlled by spectroscopy, as in the work of Sharp et al. (2006). But a broad intrinsic profile would result in a loss of information, making an unambiguous spectroscopic fit more problematic. For instance, application of the Gaussian in Figure 5 to the generic model spectra in Figure 2 would fill in the blue edge of the RRB as well as one would like.

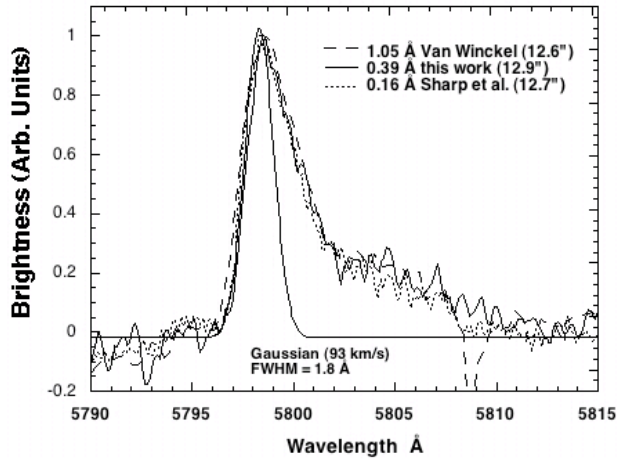

**Fig. 5** The 5799 Å band recorded at 3 different spectral resolutions at the same position along the NE spike, showing that improved resolution does not appreciably effect the characterization of the band. The blue edge could be fit by a Gaussian as shown; the red edge is presumably fit by the spectroscopic behavior of the emitter.

To address this issue we looked for spectra of known species that might inhabit the same space as the unknown emitter. Spectra of Na I and [N II] lines observed in the nebula are shown in Figure 6. Sodium is seen to have a line profile that is broader than the WIYN line-spread function at all positions in the nebula. Inside the cone walls, the line displays the "double-peaked" profile as observed by Hobbs et al. (2004) in their on-star measurements. We measure the FWHM to be about 0.6 Å, but with the caveat that sodium is not a perfect tracer of the RRBs (Schmidt and Witt 1991). It would be well worth investigating in detail the velocity behavior of sodium across the nebula with high spectral and spatial resolution.

The [N II] line displays a relatively broad profile, but is weak beyond about 3.2 arcsec. It is unclear whether [N II] tracts the RRB behavior, but the line may give an indication that a modest velocity dispersion is present in the outflow. Hobbs et al. (2004) also observe several other broad lines on the star; a few, like Na, are not simple, one-component Gaussians.

Referring again to Figure 5, the highest resolution VLT spectra show some weak substructure on the 5799 Å band, which may be narrower than 90 km/s. Therefore, we estimate that the intrinsic width is between 0.6 and 1.6 Å (30 and 80 km/s). We do not believe that this velocity dispersion is extreme. Emisson from CO (Sitko et al. 2008) and Hα (Jura et al. 1997) indicates the presence of high-velocity outflows ( >100 km/s) in the inner regions of the system. Additionally, the outflow models of Icke (2003) and Soker (2005) require relatively large gas velocities in order to blow out the nebula. Therefore, any velocity spread in the outer regions could be a remnant of the high-velocity flows from the inner regions of the system, perhaps as the eddies or vortices described by Cohen et al. (2004). In the spectra analyzed here, little difference was observed in the profiles of bands observed in the regions between the spikes as compared to those observed on the spikes (inside 7 arcsec), such as those in Figure 5. Additionally, it should be remembered that a 1.5 arcsec aperture samples a rather wide and very long column of material at any point in the nebula. In the future, spectra obtained at spatial resolutions of less that 0.3 arcsec might resolve features in the band in Figure 5 if its shape is velocity controlled. Some other lines of sight may show a compound intrinsic profile, one that has both broad and narrow components.

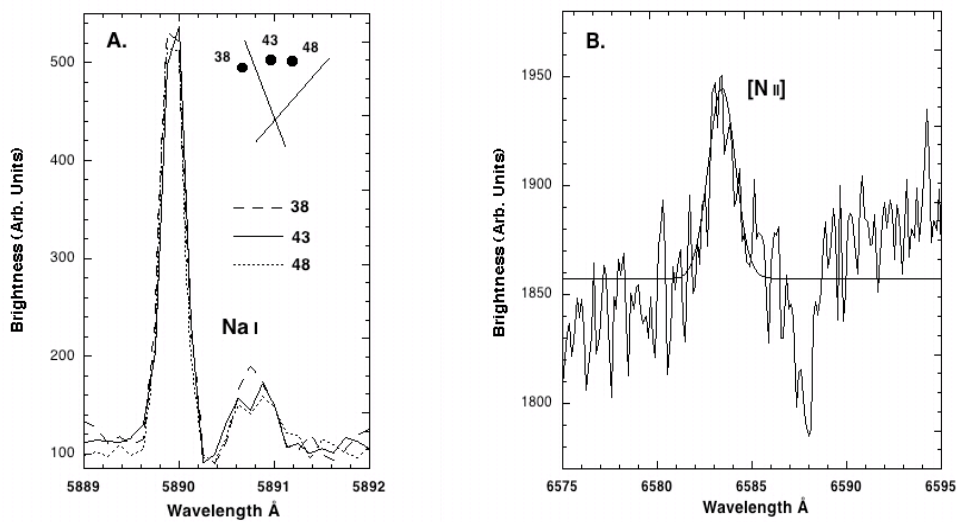

**Fig. 6** The Na D1 and [N II] line profiles in the nebula. (A) Sodium line observed in WIYN spectra at 3 positions equidistant from the central source, 10.6 arcsec. (The intense line is due to scattered city lights.) (B) Gaussian fit to the forbidden [N II] line observed at 3.2 arcsec, approximately on the NE spike; the FWHM of the Gaussian is 1.6 Å. The spectra were acquired on 27 Feb 2004 and 3 Nov 2002, respectively.

## 7 How many bands are a part of the same system?

It can be seen in several published spectra, that there is a wide array of sharp features atop the ERE (Schmidt et al. 1980; Warren-Smith et al. 1981; Van Winckel et al. 2002). Which of these bands belongs to an emission system originating from the same carrier is an open question. We think it is reasonable to assume that bands of congruent shape may belong to the same transition, but this is not to say that bands of slightly different shapes cannot. In Figure 7, we show two of the strongest RRBs overlaid. The cores of the two bands are entirely congruent. Are the 5799 and 6614 Å bands a part of the same emission system? If they are, they and other bands must fit in a comprehensive model of the system with all bands predicted to be in that system appearing in the RRB spectrum with the expected relative intensities.

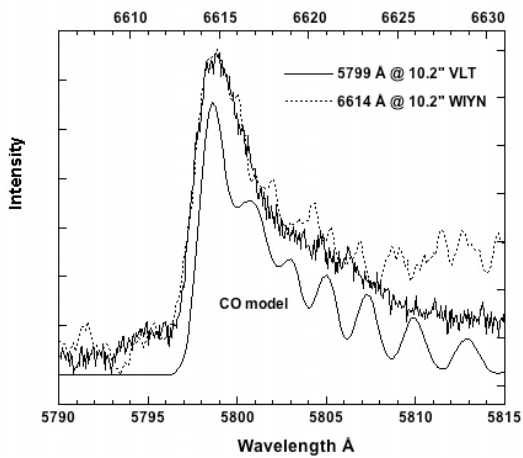

**Fig. 7** Spectra of the 5799 and 6614 Å bands overlaid in wavelength space (an overlay in velocity space would be only slightly different). The excess flux beyond 6625 Å in the WIYN spectrum is primarily due to an imprecise flat field correction; the spectrum also shows the greater noise of that data. The CO model uses $B´ = 1.4$ cm$^{-1}$; $B´´ = 1.9$ cm$^{-1}$; $T = 40$ K; and an intrinsic line width of 1.6 Å (see below).

The large number of weak RRBs may suggest the possibility of overlap. Examination of an individual 5799 Å band in Figures 1, 5 and 7 suggests there is substructure that may or may not belong to the band itself. The weak structure at 5795 and 5807 Å are cases in point. If these are from a different emission system, then some of the intensity near 5795 and 5807 Å may need to be subtracted from the 5799 Å band before modeling is attempted. We might suggest that careful analyses of other IFU spectra having greater spatial resolution may be useful in defining these relationships.

In Figure 8 we show our new identification of a $C_2$ Swan band in the Red Rectangle from VLT spectra. Also shown, for comparison, is a WIYN spectrum of the same band obtained during an observing campaign of Comet Hale-Bopp (Glinski et al. 2001). Matches of the rotational substructure on the blue side of the bandhead would seem to confirm the assignment to $C_2$; although a much higher rotational excitation temperature is evident in the comet spectrum. The feature was noted as an unidentified line in the Red Rectangle spectrum of Van Winckel et al. (2002). Other overlapping, blue-shaded bands have been reported in the work of Scarrott et al. (1992) and Van Winckel et al. (2002); Sarre (2006b) has discussed these bands as being from the $C_2$ Swan system. Preliminary mapping of the $C_2$ indicates that it inhabits regions closer to the center of the object than the RRBs (Schmidt & Sharp, unpublished); we do not know, however, why these lines are missing in the list of molecules observed on the star by Hobbs et al. (2004). Although the 5807 Å feature discussed above does not seem to belong to the Swan system, an analysis of a family of blue-shaded bands could help in their identification.

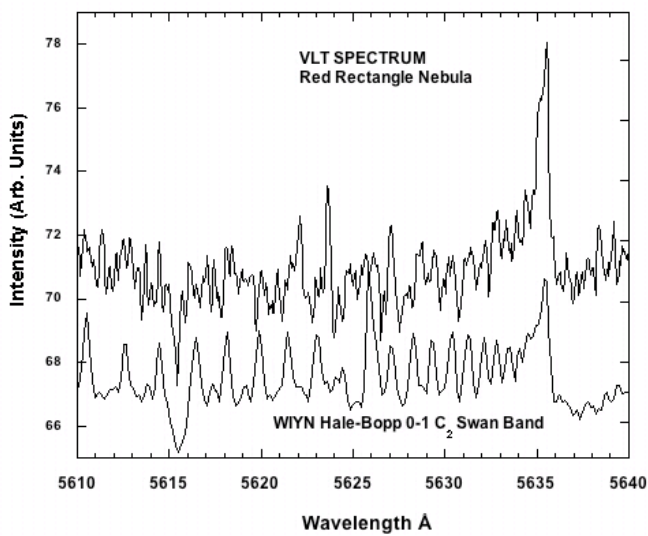

Fig. 8 VLT spectrum obtained at 5.1 arcsec on the Southeast spike shown together with a WIYN-Hydra spectrum of the clearly identifiable, blue-shaded, 0-1 bandhead of the $C_2$ Swan ($d^3\Pi_g \rightarrow a^3\Pi_u$) system.

## 8  Is there evidence of optical CO emission bands?

It has been reported earlier that carbon monoxide gives rise to several strong emission band systems in the ultraviolet (Sitko 1983; Glinski et al. 1997a). Additionally, absorption is seen into high vibrational states of both the $A^1\Pi$ state and several triplet states (Sitko et al. 2008). Phosphorescence from the Cameron bands is also observed (Glinski et al. 1996; Yan et al. 2000). This suggests that there is a cascade of emission transitions that may yield visible photons (Glinski et al. 1997a; Gadipadi and Kalb 1998; Gadipadi 2002; Sitko et al. 2008). Bennett et al. (1999) have suggested that perhaps all or some of the ERE may be due to emission from CO. It is disappointing, however, that expected transitions from known optical emission systems are not readily apparent. A search using tables of band-head positions of the Asundi or Triplet bands does not yield a good set of matches to many RRBs (Krupenie 1966; Tillford and Simmons 1971).

We noticed that the frequency difference between the strong RRBs in Figure 6 is 2128 cm$^{-1}$; this is very similar to a vibrational frequency in an electronic excited state of CO (Huber and Herzberg 1979). As an exploration, we constructed a preliminary model for a hypothetical CO transition that might appear in the visible like the bands in Figure 7. The models represents the simplest case of a $^1\Sigma - {}^1\Sigma$ transition. The rotational constants we used are consistent with a lower electronic state that has a fundamental vibrational frequency of about 2200 cm$^{-1}$ and an upper electronic state that has a fundamental vibrational frequency of about 1100 cm$^{-1}$. One could speculate that these electronic states could correspond to be two states, lying at energies from 80,000 to 110,000 cm$^{-1}$ (10 to 13 eV) (Huber and Herzberg 1979). The ultraviolet spectra show that there is absorption into the $A^1\Pi$ state to at least 10 eV (Sitko et al. 2008). The model spectrum in Figure 7 would seem to present a problem for the CO hypothesis. Even when the intrinsic line width is made as large as defensible to fill in the blue edge, prominent rotational structure is still predicted.

## 9  Conclusions

We, and others, are interested in spectroscopic models that might assign groups of RRBs to particular molecules. At present, it is possible to fit the spectra outside of about 8 arcsec by several molecules, medium to large, under reasonable conditions. In this paper, we do not propose a solution to the problem, but aim to present a cautious evaluation of the current observational constraints for future analysis. The following are the pertinent issues, which should be addressed:

1.) Absorption effects are strongly suggested, but not proven. These effects could yield valuable clues to the identity of the emitter. Better characterization of this absorption would provide additional constraint on the assignment. At this juncture, the carrier would appear to be a good absorber and an efficient fluorescer. If an absorption spectrum could be deconvolved from the emission spectra, it could provide a stronger test for the models.

2.) A properly defined line profile is vital to the modeling, but the relatively low spatial resolution of the ground-based observations makes it problematic to deduce an intrinsic line profile. It appears the intrinsic width is between 0.6 and 1.6Å. The overall appearance of the bands may be distorted by the sampling of mixed environments along the lines of sight in a wide aperture, highlighting the need for space telescope observations.

Future observation campaigns should look to correlate the RRBs to known spectra, e.g., with CO and $C_2$; correlation of velocity spreads would be quite valuable. We think it would be of greatest interest to map the UV CO emission and some of the atomic emission lines, such as the Na I and [N II] lines, in regions beyond the inner few arcseconds. Additionally, where as the ERE is ubiquitous, the only other possible observation of the sharp RRBs is in the deep minimum spectrum of the R CrB star V 854 Cen (Rao & Lambert 1993). It may be useful to revisit this object (as a target of opportunity) to examine the time-evolution of the emission features during the occultation event.

We emphasize that serious ambiguities still exist in the RRB data. Critical analysis leads us to think that the identity of the carrier of the RRBs is still only partially constrained by the spectra. We hope that these insights will serve as an informed guide for future observations and modeling studies.

This paper is based partly on presentations and discussions, which occurred at the colloquium entitled "The Red Rectangle," at University of Virginia, Charlottesville during May 2006.